\documentclass[10pt,letterpaper]{article}
\usepackage{opex3}

\usepackage{here}
\usepackage{graphicx}

\newcommand\pictc[5]{\begin{figure}[t]
            \centerline{\vspace{0mm}
\includegraphics[width=#1\columnwidth,height=0.7\textheight,keepaspectratio]{#3}}
            \protect\caption{\protect\label{fig:#4} #5}
                    \end{figure}            }

\newcommand\pict[4][1]{\pictc{#1}{!tb}{#2}{#3}{#4}}
\newcommand\rpict[1]{\ref{fig:#1}}

\newcommand\leqt[1]{\protect\label{eq:#1}}
\newcommand\reqtn[1]{\ref{eq:#1}}
\newcommand\reqt[1]{(\reqtn{#1})}

\newcommand\lsect[1]{\protect\label{sect:#1}}
\newcommand\rsect[1]{\ref{sect:#1}}

\newcounter{Fig}

\begin{document}

\title{Crossover from self-defocusing to discrete trapping in nonlinear waveguide arrays}

\author{Michal Matuszewski$^{1,2}$, Christian~R. Rosberg$^1$, Dragomir~N. Neshev$^1$, Andrey~A. Sukhorukov$^1$, Arnan Mitchell$^4$, Marek Trippenbach$^{2,3}$, Michael~W. Austin$^4$, Wieslaw Krolikowski$^1$, and Yuri~S. Kivshar$^1$}
\address{$^1$Nonlinear Physics Centre and Laser Physics Centre, Centre for Ultrahigh-bandwidth Devices for Optical Systems (CUDOS), Research School of Physical Sciences and Engineering, The Australian National University, Canberra ACT 0200, Australia\\
$^2$Institute for Theoretical Physics, Warsaw University, Ho\.{z}a 69, PL-00-681 Warsaw, Poland\\
$^3$Soltan Institute for Nuclear Studies, Ho\.{z}a 69, PL-00-681 Warsaw, Poland\\
$^4$School of Electrical and Computer Systems Engineering, RMIT University, Melbourne}
\email{dnn124@rsphysse.anu.edu.au}
\homepage{http://www.rsphysse.anu.edu.au/nonlinear} 

\begin{abstract}
We predict a sharp crossover from nonlinear self-defocusing to discrete self-trapping of a narrow Gaussian beam with the increase of the refractive index contrast in a periodic photonic lattice.
We demonstrate experimentally nonlinear discrete localization of light with defocusing nonlinearity by single site excitation in LiNbO$_3$ waveguide arrays.
\end{abstract}


\section{Introduction}

The growing interest in the study of nonlinear periodic structures is motivated by the fact that novel physical phenomena can be observed due to the interplay between nonlinearity and periodicity~\cite{Christodoulides:2003-817:NAT}. The presence of periodically varying refractive index in the medium results in a bandgap structure of the transmission spectrum. This subsequently affects the propagation of optical beams, which is determined by the dispersion curves of different transmission bands and forbidden gaps.
By engineering a periodic structure, it becomes possible to manage the strength and type of wave diffraction~\cite{Eisenberg:2000-1863:PRL}, and therefore control the self-action of light in nonlinear media~\cite{Morandotti:2001-3296:PRL}. If the refractive index decreases with light intensity due to nonlinear response of the material, the beam normally experiences broadening due to the self-defocusing. However, in periodic photonic structures, the same type of nonlinearity allows for beam localization, and this effect is analogous to nonlinear self-trapping in discrete systems~\cite{Christodoulides:2003-817:NAT}.

In this paper, we study the self-action dynamics of an initially narrow Gaussian beam propagating in a nonlinear defocusing medium with periodically modulated refractive index. We demonstrate that there is a sharp transition from self-defocusing to discrete self-trapping when the depth of index  modulation ($\Delta n$) is increased. We also observe experimentally the formation of a self-trapped state from a single-site excitation in the defocusing regime.

\section{Discrete self-trapping in photonic lattices} \lsect{discrete}

Nonlinear propagation of light in photonic structures with a periodic modulation of the optical refractive index in one transverse spatial dimension, such as waveguide arrays~\cite{Christodoulides:2003-817:NAT, Morandotti:2001-3296:PRL, Eisenberg:1998-3383:PRL, Iwanow:2004-113902:PRL, Chen:2005-4314:OE, Fratalocchi:2004-1530:OL} or optically-induced photonic lattices~\cite{Fleischer:2003-23902:PRL, Fleischer:2003-147:NAT, Neshev:2003-710:OL, Martin:2004-123902:PRL}, is commonly described by the so-called tight-binding approximation~\cite{Ashcroft:1984:SolidState, Christodoulides:1988-794:OL},
\begin{equation} \leqt{discrete}
   i \frac{d a_n}{dz} + \beta a_n + C ( a_{n-1} + a_{n+1} ) + \gamma |a_n|^2 a_n = 0,
\end{equation}
where $z$ is the propagation coordinate, $a_n(z)$ is the mode amplitude in the $n-$th waveguide [Fig.~\rpict{lattice}(a)], and $\beta$ is the propagation constant. Coefficient $C$ stands for the nearest-neighbor coupling between the waveguides, and the last term in Eq.~\reqt{discrete} accounts for mode detuning through the intensity-dependent change of the refractive index.

In the case of self-focusing nonlinearity, Eq.~\reqt{discrete} supports the so-called {\em unstaggered discrete solitons}~\cite{Christodoulides:2003-817:NAT, Christodoulides:1988-794:OL}. These are strongly localized states where the light is confined to a few waveguides, and the amplitudes of the modes of neighboring waveguides are in-phase. Importantly, defocusing nonlinearity can also support discrete self-trapping of light, but in the form of {\em staggered solitons} when the amplitude of the modes in neighboring waveguides is out of phase~\cite{Kivshar:1993-1147:OL}. This can be understood by noting that the form of Eq.~\reqt{discrete} remains unchanged after the transformation: $a_n \rightarrow (-1)^n {a}_n^\ast$ and $\gamma \rightarrow -\gamma$. It follows that the beam dynamics in the framework of Eq.~\reqt{discrete} is fully equivalent for positive ($\gamma>0$) and negative ($\gamma<0$) nonlinearities, with the only difference being in the phase structure, provided only a single site is excited at the input, i.e. $a_n(z=0)=0$ for $n\ne0$. In particular, when the input intensity is high enough, a discrete soliton should form for either type of nonlinear response.

\pict[0.9]{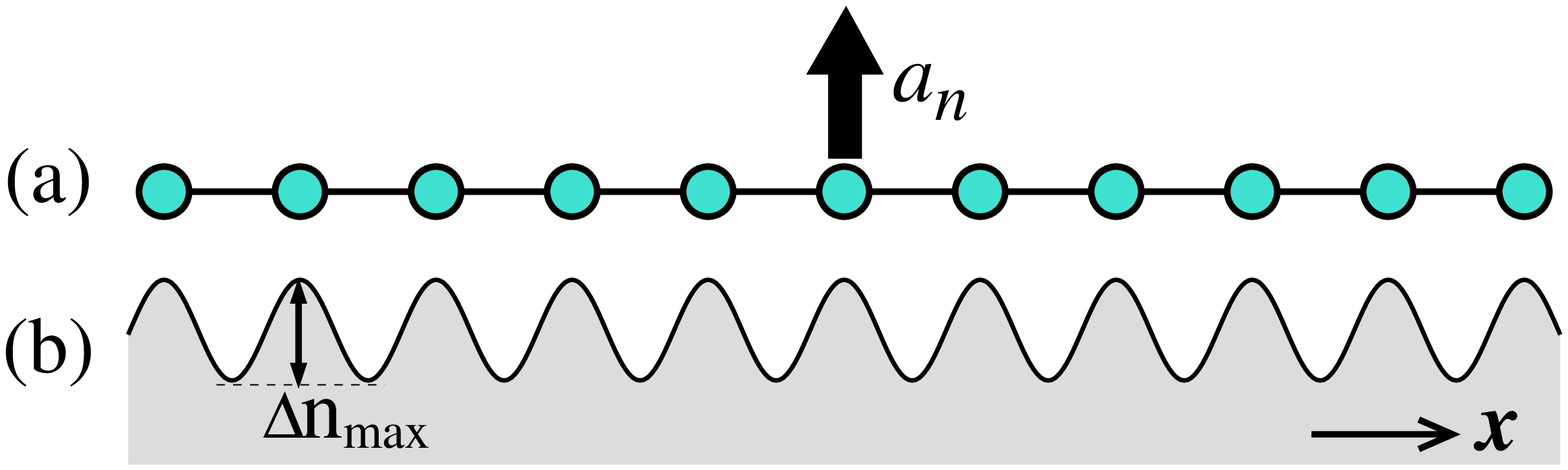}{lattice}{
(a)~Schematic of a discrete lattice with nearest-neighbor coupling.
(b)~Refractive index profile of coupled LiNbO$_3$ waveguides.}

\section{Nonlinear localization in continuous periodic photonic structures}

The tight-binding model~\reqt{discrete} predicting a universal self-trapping scenario is valid when the optical field can be represented as a superposition of weakly overlapping modes of the individual lattice sites. Such condition is generally satisfied when the refractive index contrast in a periodic structure is sufficiently large~\cite{Alfimov:2002-46608:PRE}. However, in the case of defocusing nonlinearity, the beam dynamics changes dramatically for smaller index contrast. Indeed, if the lattice modulation is absent or very weak, the beam will only experience nonlinearly enhanced defocusing instead of self-trapping. In order to study the crossover between beam self-defocusing and discrete self-trapping, we model the beam propagation using a normalized continuous nonlinear Schr\"odinger equation~\cite{Kivshar:2003:OpticalSolitons} for the slowly varying field envelope $E(x,z)$,
\begin{equation} \leqt{nls}
   i \frac{\partial E}{\partial z}
   + D \frac{\partial^2 E}{\partial x^2}
   + {\cal F}(|E|^2) E
   + \rho\; \Delta n(x)\; E = 0,
\end{equation}
where $D=z_s \lambda/(4 \pi n_0 x_s^2)$ is the diffraction coefficient, $\rho=2 \pi z_s / \lambda$, $x$ is normalized to $x_s$, $z$ is normalized to $z_s$. We choose the parameters to match the conditions of our experiments described below. The linear refractive index change $\Delta n$ is taken as $\Delta n(x) = \xi \sum_n \exp[-(x-n\;d)^2 / w^2]$, where $\xi$ defines the modulation depth. We take the values of waveguide width $w=12 \mu$m and the waveguide spacing $d=19 \mu$m, and then the corresponding refractive index contrast is $\Delta n_{max}=0.442\; \xi$. We note that this index profile [Fig.~\rpict{lattice}(b)] is defined by the experimental realization, however we have verified that our conclusions are valid for different lattice profiles. Other parameters are: $\lambda = 0.532 \mu$m, $n_0=2.234$, $x_s=1 \mu$m, $z_s=1$mm, and ${\cal F}(I)=1.5 (1 + I)^{-1}$ for photovoltaic defocusing nonlinearity~\cite{Valley:1994-4457:PRA}.

\pict[1]{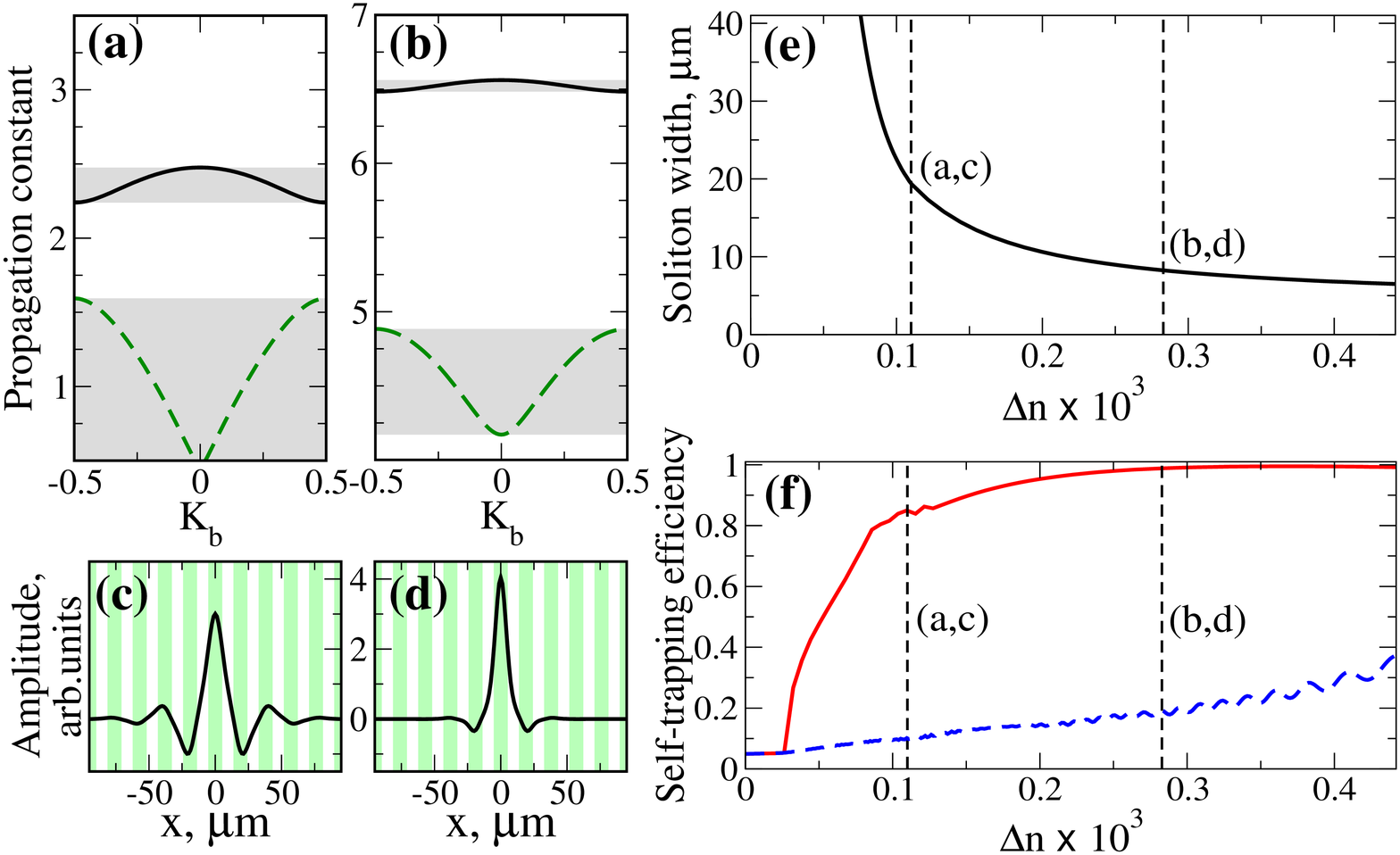}{theory}{
(a,b)~Bandgap spectrum of linear waves for index contrasts of $1.1\times10^{-4}$ and $2.8\times10^{-4}$, respectively.
(c,d)~Profiles of staggered gap solitons having the minimum width for cases (a,b), respectively. Shading marks index maxima.
(e)~Minimum width of the gap soliton vs. the refractive index contrast;
(f)~Efficiency of beam self-trapping calculated as the power fraction remaining in the 20 central waveguides at the output for an optimized input power (solid) compared with linear diffraction (dashed) vs. the refractive index contrast.
}

The periodic modulation of the refractive index results in the formation of a bandgap structure for the wave-vector components, as the light propagates along the waveguides. In Figs.~\rpict{theory}(a,b) we present the calculated bandgap spectrum (see Ref.~\cite{Sukhorukov:2004-93901:PRL} for details of such calculations) for the waveguide array shown in Fig.~\rpict{lattice}(b) for two values of the refractive index contrast $\Delta n_{max}$. In the case of defocusing nonlinearity, solitons can form at the bottom edge of the first band, and their propagation constant is shifted deeper into the gap for larger intensities. This scenario is predicted correctly by the tight-binding model~\reqt{discrete}, however this model does not account for the existence of the second band, that defines the gap extent. On the other hand, the gap size limits the minimum width of self-trapped beams, see examples in Figs.~\rpict{theory}(c,d). The width of the gap is smaller for weaker refractive index contrast and increases for larger $\Delta n_{max}$. The calculated minimal width of the gap soliton, $W = 3 \int |x| |\psi|^2 dx / \int |\psi|^2 dx$, is plotted in Fig.~\rpict{theory}(e) vs. the refractive index contrast. In the case of small refractive index contrast and a narrow band gap, the narrowest soliton spans over several waveguides. The situation changes when the contrast of index modulation increases, and the narrowest soliton is localized at a single waveguide.

\section{Crossover from self-defocusing to discrete self-trapping}

We study the crossover to discrete self-trapping by modeling the dynamics of an input Gaussian beam, which width is equal to the size of a single waveguide. In Fig.~\rpict{theory}(f) we plot the relative power that remains in the central section of the array containing 20 waveguides 
(implemented with absorbing boundary conditions) after the propagation over a distance of many diffraction lengths (1000mm) vs. the index contrast of the lattice.
We investigate the effect of nonlinearity (solid line) on beam dynamics with respect to linear diffraction (dashed line).
We optimized the power of the input beam in order to maximize the power fraction which remains in the central section of the array. Our results demonstrate that for low index contrast the optimization of the input power for nonlinear propagation simply matches the linear limit, and the two curves coincide in Fig.~\rpict{theory}(f). In this regime, the nonlinear self-action results in increased beam spreading due to self-defocusing.
As the refractive index increases, there appears a bifurcation from the linear regime. This corresponds to the formation of initially broad gap solitons, in agreement with the limitation on the soliton width shown in Fig.~\rpict{theory}(e).
As the minimum width of the soliton decreases and approaches that of a single guide for larger $\Delta n_{max}$, self-trapping with almost 100\% efficiency becomes possible, as predicted by the discrete model~\reqt{discrete}. This feature represents a transition from self-defocusing to discrete self-trapping as the index contrast exceeds a certain threshold.

\pict{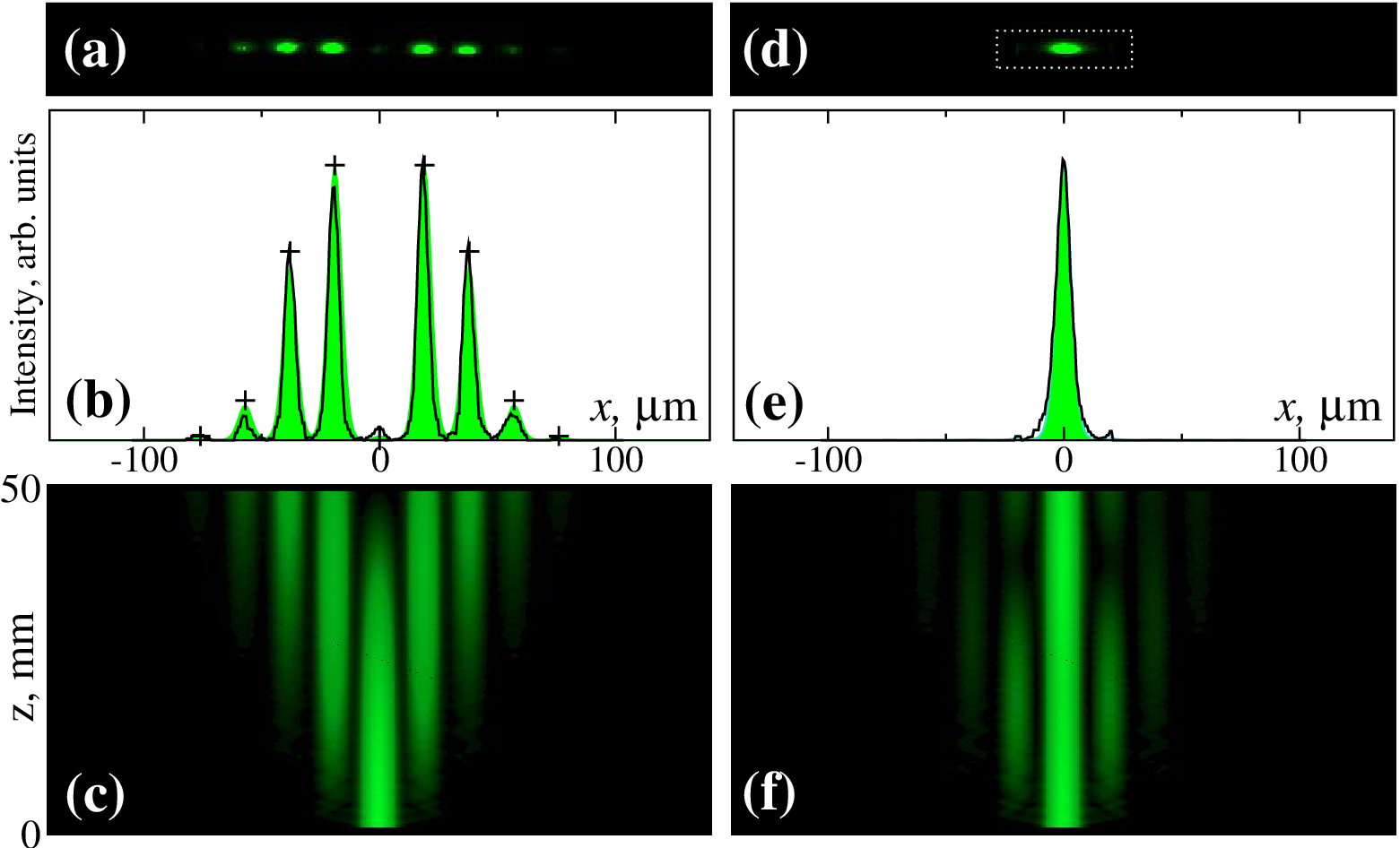}{experiment}{
(a)~Intensity distribution at the output facet of the array for linear propagation at low laser power (10~nW). (b)~Corresponding intensity profiles: Solid -- experimental measurement; shading -- numerical solution of the full model [Eq.~\reqt{nls}]; crosses -- lattice site amplitudes calculated from the discrete model [Eq.~\reqt{discrete}].
(c)~Evolution of the beam intensity along the sample for a low input power simulated with Eq.~\reqt{nls}. (d-f)~Same as (a-c) for nonlinear propagation at high laser power (1~mW).}

\section{Experimental observation of single-site self-trapping with defocusing nonlinearity}

In previous experimental studies of soliton formation in permanent~\cite{Iwanow:2004-113902:PRL, Chen:2005-4314:OE} and optically-induced~\cite{Fleischer:2003-23902:PRL, Fleischer:2003-147:NAT} periodic structures with defocusing nonlinearity, the input excitation was specially prepared to reflect the phase structure of the staggered solitons, where the neighboring sites are out-of-phase [see Figs.~\rpict{theory}(c,d)]. This was achieved either by illuminating the structure with a single beam inclined at the Bragg angle~\cite{Iwanow:2004-113902:PRL, Fleischer:2003-23902:PRL, Fleischer:2003-147:NAT} or by phase modulation of the initial beam~\cite{Chen:2005-4314:OE}. However, observation of a soliton localization at a single waveguide was not reported.

In order to study experimentally the discrete self-trapping at a single site supported by defocusing nonlinearity, we designed an array of closely spaced optical waveguides produced by Titanium indiffusion into a mono-crystal lithium niobate wafer. The waveguide array has a period of 19$\mu$m and refractive index contrast $\Delta n_{max}=2.8\times 10^{-4}$, which was chosen above the threshold for a crossover to discrete self-trapping predicted in numerical simulations [Fig.~\rpict{theory}(f)]. In the fabrication process, 100{\AA} of Ti was deposited on the X-cut LiNbO$_3$ using electron beam evaporation. The Ti layer was then photolithographically patterned and etched in a buffered hydrofluoric acid solution. The diffusion was conducted at 1050$^\circ$C for 3 hours in a wet oxygen environment. The waveguides were verified as single mode
using a prism coupling technique. The array was then diced to a total length of 5cm and both facets were mechanically polished.

The LiNbO$_3$ sample exhibits a strong photovoltaic effect which leads to the negative (self-defocusing) nonlinear response to laser beams at a photosensitive wavelength. In our experiments, we tightly focused an extraordinary polarized laser beam from a cw Nd:YVO$_4$ laser into a single guide of the array by a microscope objective ($\times20$). The input and output facets were monitored by two CCD cameras. The array was externally illuminated with white light in order to reduce the nonlinear response time of the photovoltaic material to less than a minute. At low laser power ($\sim$10~nW) the propagating beam experienced typical discrete diffraction~\cite{Christodoulides:2003-817:NAT}, where at the array output most of the laser power was transferred into the neighboring waveguides and there was almost no light in the central guide [Fig.~\rpict{experiment}(a)]. The resulting intensity profile shown in Fig.~\rpict{experiment}(b, solid line) matches well the results of numerical simulations performed using the discrete model [Eq.~\reqt{discrete}] (crosses) and the full model with periodic index modulation [Eq.~\reqt{nls}] (shading). The corresponding calculated propagation inside the array is depicted in Fig.~\rpict{experiment}(c).

\pict{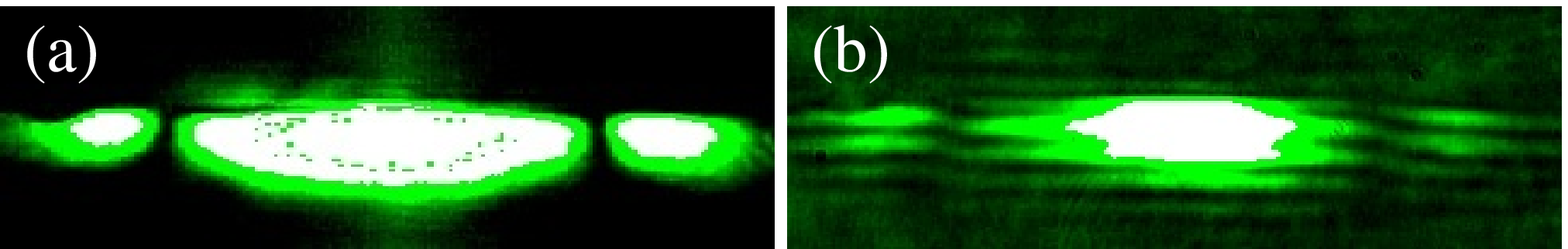}{zoom}{(a) Saturated camera image of the single channel localized state at the output of the array, zoomed at the dashed rectangle in Fig.~\rpict{experiment}(d).
(b)~Interferogram confirming the staggered phase structure of the output beam.}

When the laser power is increased (1~mW) the defocusing photovoltaic nonlinearity leads to strong beam localization at a single waveguide [Fig.~\rpict{experiment}(d-f)], in a similar way as in focusing waveguide arrays~\cite{Christodoulides:2003-817:NAT}, due to the universal nature of discrete self-trapping as discussed in Sec.~\rsect{discrete}.

In order to confirm that the nonlinear state is indeed localized inside the Bragg reflection gap, and not in the total internal reflection gap, it is important to verify its staggered phase structure. For this purpose we allowed for saturation of the camera in order to detect the small but nonzero amount of light in the neighboring waveguides. The corresponding images are shown in Fig.~\rpict{zoom} where the two neighboring satellites are clearly visible. They are separated from the central waveguide by zero-intensity lines, which is an indication of their out-of-phase structure. To confirm the staggered phase of the localized state we
interfere the output with an inclined broad reference beam. The corresponding interferogram is shown in Fig.~\rpict{zoom}(b), where a $\pi$ shift of the interference fringes at the zero intensity lines is clearly observed.

\section{Conclusions}

We have studied the crossover between the beam self-defocusing and discrete self-trapping in waveguide arrays with defocusing nonlinearity. We have shown that an abrupt transition is observed in the beam dynamics with the increase of the refractive index. For a small index contrast, the beam experiences enhanced spatial spreading at higher input powers. However, when the contrast exceeds a critical value, beam self-trapping associated with the formation of staggered gap solitons becomes possible.
We have demonstrated experimentally the generation, by single waveguide excitation, of strongly localized staggered states supported by defocusing nonlinearity in waveguide arrays, where the refractive index modulation was engineered to exceed the crossover threshold.

\section*{Acknowledgements}

This work was partially supported by the Australian Research Council. 
M. M. acknowledge support from the KBN grant 2P03 B4325, M.T. was supported by the Polish Ministry of Scientific Research and Information Technology under grant PBZ MIN-008/P03/2003.

\end{document}